\documentclass[reprint,preprintnumbers,amsmath,amssymb,nofootinbib,aps,prx]{revtex4-1}

\usepackage{slashed}
\usepackage{amsmath}
\usepackage{amssymb}
\usepackage{amsthm}
\usepackage{hyperref}
\usepackage{indentfirst}
\usepackage{psfrag}
\usepackage{graphicx}
\usepackage{cleveref}
\usepackage[utf8]{inputenc}

\hypersetup{colorlinks=true, linkcolor=blue, urlcolor=blue, citecolor=blue, linktocpage=true}

\usepackage{xifthen}
\newcommand*\diff{\mathrm{d}} 
\newcommand*\ldiff[2][]{ \ifthenelse{\isempty{#1}}{ \diff #2}{\diff^#1#2} \,} 
\let\limitint\int 
\renewcommand{\int}{\limitint \!} 

\def\e{{\rm e}}

\newcommand{\bseq}{\begin{subequations}}
\newcommand{\eseq}{\end{subequations}}

\renewcommand{\Im}{\mathop{\rm Im}\nolimits}
\renewcommand{\Re}{\mathop{\rm Re}\nolimits}

\newcommand{\be}{\begin{equation}}
\newcommand{\ee}{\end{equation}}
\newcommand{\beqa}{\begin{eqnarray}}
\newcommand{\eeqa}{\end{eqnarray}}

\newcommand\vf{\varphi}

\renewcommand{\L}{\mathcal{L}}
\renewcommand{\P}{\mathcal{P}}

\begin{document}

\vspace{-3.0cm}
\begin{flushright}
{\small FTPI-MINN-22-22,  UMN-TH-4131/22} 
\end{flushright}
\vspace{0.5cm}

\title{ Out of this world neutrino oscillations }

\author{Tony Gherghetta$^1$}
\email{tgher@umn.edu}

\author{Andrey Shkerin$^2$}
\email{ashkerin@umn.edu}

\affiliation{$^1$School of Physics and Astronomy, University of Minnesota, Minneapolis, Minnesota 55455, USA\\$^2$William I. Fine Theoretical Physics Institute, School of Physics and Astronomy, \\ University of Minnesota, Minneapolis, Minnesota 55455, USA }

\begin{abstract}

We study how vacuum neutrino oscillations can be affected by a causal, nonlinear and state-dependent modification of quantum field theory that may be interpreted using the many-worlds formulation of quantum mechanics.
The effect is induced by a Higgs-neutrino Yukawa interaction that causes a nonlinear interference between the neutrino mass eigenstates. 
This leads to a tiny change in the oscillation pattern of light, active neutrinos.
At large baselines where the oscillations disappear, the nonlinear effect is also suppressed and does not source correlations between the mass eigenstates once they are entangled with the environment.
Our example provides a way to compute effects of nonlinear quantum mechanics and field theory that may probe the possible physical reality of many worlds.

\end{abstract}

\maketitle

\section{Introduction}
\label{sec:intro}

Quantum mechanics is one of the pillars of modern physics. It provides an unparalleled description of the microscopic world that has been experimentally tested to an unprecedented precision. 
Nonetheless, lurking underneath this remarkable success is the tantalising question of whether quantum mechanics admits a generalisation extending from its core principles.
The answer to this question is not so straightforward because not only must any extension preserve the probabilistic interpretation of squared amplitudes or the Hilbert space structure of states in a quantum system, but it must also be consistently embedded into the framework of quantum field theory.

A possible way to modify the linear evolution of a quantum system is to introduce a small nonlinear term in the underlying Schr\"odinger equation \cite{Kibble:1978vm,Weinberg:1989us,Kaplan:2021qpv}.
In the context of generalising quantum mechanics, this modification is not thought of as induced by the interaction of the system with the environment; rather it is assumed to be a fundamental property of the system itself.
Nonlinear quantum mechanics relinquishes the principle of linear time evolution and requires an adjustment of certain concepts such as measurement \cite{Weinberg:1989us,Kaplan:2021qpv}.
In exchange it offers a plethora of possible tests spanning from measurements of atomic energy level splittings to the cosmological history of the universe \cite{Weinberg:1989cm,Weinberg:2016uml,Kaplan:2021qpv,Raizen:2022xlv,Polkovnikov:2022hkg,Broz:2022aea}.

Among possible nonlinear modifications of quantum mechanics, the proposal of Ref.~\cite{Kaplan:2021qpv} stands out because it admits a straightforward reformulation in field-theoretic terms.
At the Lagrangian level, the nonlinear correction is introduced in the form of an interaction between a field and the expectation value of some bosonic operator.
For example, for a fermion field $\Psi$ coupled to an operator $\cal O$, one can schematically write
\begin{equation}\label{NLterm}
\mathcal{O}\bar{\Psi}\Psi \mapsto \epsilon \langle\Phi|\hat{\mathcal{O}}|\Phi\rangle\bar{\Psi}\Psi \;,
\end{equation}
where $|\Phi\rangle$ represents the combined $\Psi, {\cal O}$ system and $\epsilon$ is a small dimensionless parameter.
This causes a nonlinear evolution for the field $\Psi$ (or the state vector $|\Phi\rangle$), while retaining causality and unitarity of the original theory \cite{Kaplan:2021qpv}, unlike the other proposals \cite{Kibble:1978vm,Weinberg:1989us} reviewed in Refs.~\cite{GISIN19901,Polchinski:1990py,Kaplan:2021qpv}.

Intriguingly, nonlinear generalisations of quantum mechanics have a sensible physical interpretation \cite{Polchinski:1990py,Kaplan:2021qpv}, which is revealed using the Everett (or many-world) formulation of quantum mechanics~\cite{Everett:1957hd} (see also \cite{dewitt2015many}).
Consider a system consisting of a coherent superposition $|\Phi\rangle$ of states (with respect to some basis) that interacts with the environment.
For example, each state can be a Gaussian wave packet describing a free particle.
In the standard framework, as the system freely evolves, the states interfere with each other while gradually losing coherence as $|\Phi\rangle$ becomes entangled with the environment. Eventually, the states decohere completely and, in Everett's interpretation, become part of distinct ``quasiclassical histories'' that no longer interfere with each other.
If the nonlinear term (\ref{NLterm}) is fundamentally present in the theory, then different states in the superposition  $|\Phi\rangle$ become coupled to each other.
This can change the interference pattern if the states constituting $|\Phi\rangle$ are coherent; at late times, when the system branches and quasiclassical histories form, the nonlinear term can also maintain some degree of coherence between these histories. 
In other words, different branches of the system (consisting of a few particles or, possibly, encompassing the entire universe) can, in principle, feel each other's presence via the nonlinear term.

Several experimental setups have been proposed \cite{Raizen:2022xlv,Polkovnikov:2022hkg,Broz:2022aea} to test the modification of Ref.~\cite{Kaplan:2021qpv}.
They deal with non-relativistic systems and employ the electromagnetic interaction as a basis for the nonlinear extension, i.e., the operator $\mathcal{O}$ in \cref{NLterm} represents the photon field.
In this paper, we consider for the first time an ultra-relativistic system to test the predictions of nonlinear quantum mechanics.
Furthermore, instead of the photon field, we employ the massive, scalar (Higgs) field as a messenger linking the states of the system.
Specifically, we study how the modification of the form (\ref{NLterm}) affects vacuum oscillations of the neutrino.
One can expect that the correction to the free neutrino propagation caused by the nonlinear modification accumulates over the propagation distance.
This could make the nonlinear effect more pronounced than in systems containing bound states.
Besides, a weakly interacting neutrino maintains coherence over macroscopic distances and therefore 
the neutrino is a unique system for testing modifications of quantum mechanics. 
In particular, this allows us to place the first bound on the nonlinearity parameter $\epsilon$ resulting from the Higgs-neutrino Yukawa interaction.

We consider the light, active neutrino produced in flavour $a$ which propagates some distance (baseline) $L$ before it is detected in flavour $b$.
At the moment of production, the neutrino is a superposition of $N_f$ mass eigenstates, where $N_f$ is the number of active neutrino flavours. 
We assume that these mass eigenstates can be modeled by Gaussian wave packets localised in phase space, which propagate in a given direction.
Initially, the wave packets interfere with each other giving rise to the standard picture of neutrino oscillations. These oscillations will end when, due to the difference in their group velocities, the wave packets no longer overlap. 
We assume that the neutrino interaction with the environment 
is sufficiently weak and does not disturb the neutrino propagation.  
However, the interaction with the environment may cause decoherence when the wave packets in the superposition have large spatial separation. 
In other words,  according to the many-worlds interpretation,  the system branches into distinct quasiclassical histories,  each containing one mass eigenstate.\:\footnote{A further branching of the system occurs as a result of neutrino detection. 
Correlations between these branches, induced by the nonlinear correction, can also affect the measurement outcome.
We do not consider this effect.}
The nonlinear correction to the neutrino propagation can affect the interference between the wave packets when they are still coherent, and can also prevent the quasiclassical histories from completely decohering at late times.
Our goal is to study these effects by computing the correction to the oscillation probability as a function of $L$,  to first order in the small nonlinearity parameter $\epsilon$.

The paper is organised as follows.
In Sec.~\ref{sec:vac} we review the wave packet treatment of linear neutrino oscillations in the vacuum.
In Sec.~\ref{sec:nonlin} we first provide a general framework for computing the nonlinear correction to the neutrino evolution amplitude and oscillation probability.
The equation of motion for the neutrino wavefunction is written in the form of a Schr\"odinger equation and we show that the nonlinear correction to the wavefunction satisfies an inhomogeneous Schr\"odinger equation.
To determine the inhomogeneous term, we consider Majorana neutrinos and apply the prescription (\ref{NLterm}) to the Higgs-neutrino interaction arising from the Weinberg operator after integrating out heavy Majorana states.
We then compute the correction to the oscillation probability induced by this extension, which is the main result of the paper.
We discuss particular cases and possible implications of our result in Sec.~\ref{sec:phys}.
Finally, our concluding remarks are given in Sec.~\ref{sec:disc}.
Several appendices contain details of the calculations and further discussion. 
We use natural units $\hbar=c=1$.

\section{Vacuum neutrino oscillations}
\label{sec:vac}

We first review neutrino oscillations in vacuum.
For illustrative purposes and to make contact with the standard treatment of neutrino oscillations, we adopt the notation--- time-evolving state vectors and operators acting on them---from quantum mechanics.
The more accurate, field-theoretic description will only be needed to derive the explicit form of the nonlinear correction to the neutrino propagation.

We assume that the initial neutrino state is described by a superposition of wave packets propagating in a particular direction, which we choose to be along the $z$ axis.
Let ${\vec x}=(x,y,z)=(0,0,z_{\rm p})$ and $t=t_{\rm p}$ be the position and time of production, respectively, of a particular neutrino flavour state $|\nu^{({\rm p})}_a(t_p)\rangle$.
The normalised flavour neutrino state at $t> t_{\rm p}$ is then given by
\begin{equation}\label{WavePacket}
    \vert \nu^{({\rm p},0)}_a (t)\rangle 
    = \sum_{k=1}^{N_f} \bar{V}_{a k} \int\frac{\diff^3\vec{p}}{(2\pi)^{3/2}} 
    f_{\rm p}(\vec{p})\:\e^{-iE_k(t_{\rm p}-t)} \left\vert \nu_k(\vec{p}) \right\rangle \;.
\end{equation}
Here, $V_{i k}$ is the PMNS mixing matrix, the bar notation denotes complex conjugation, $E_k=\sqrt{\vec{p}^2+m_k^2}$ and $\nu_k(\vec{p})$ is the $k$-th mass eigenstate.
The wave packet profile is assumed to be the Gaussian
\begin{equation}\label{GaussianPacket}
    f_{\rm p}(\vec{p}) = \left( \frac{2\pi }{\sigma^2} \right)^{\frac{3}{4}} \exp \left[- \frac{p_x^2}{4\sigma^2}- \frac{p_y^2}{4\sigma^2} -\right. \left.\frac{(p_z-p_{\rm p})^2}{4\sigma^2}+ip_z z_{\rm p} \right],
\end{equation}
where, for simplicity, the momentum uncertainty $\sigma$ is chosen to be equal for all momentum components. 
The wave packets in (\ref{WavePacket}) move with average momentum $p_{\rm p}$ up until they are detected.
Let ${\vec x}=(0,0,z_{\rm d})$ and $t=t_{\rm d}$ be the position and time of detection, respectively.
The state vector of the detector is
\begin{equation}
    \vert\nu^{({\rm d})}_b \rangle 
    = \sum_{k=1}^{N_f} \bar{V}_{b k} \int\frac{\diff^3 \vec{p}}{(2\pi)^{3/2}} 
    f_{\rm d}(\vec{p}) \left\vert \nu_k(\vec{p}) \right\rangle \;,
\end{equation}
where $f_{\rm d}(\vec{p})$ is given by \cref{GaussianPacket} with the replacement $(p_{\rm p},z_{\rm p})\mapsto (p_{\rm d},z_{\rm d})$.
The amplitude for producing a flavour $a$ neutrino and detecting a flavour $b$ neutrino at time $t_{\rm d}$ is then given by $\mathcal{A}^{(0)}_{ab}  = \langle\nu^{({\rm d})}_b | \nu^{({\rm p},0)}_a(t_{\rm d})\rangle = \langle\nu^{({\rm d})}_b | U_0(t_{\rm p},t_{\rm d})\nu^{({\rm p})}_a(t_{\rm p})\rangle$, where $U_0$ is the standard linear evolution operator. 
Projecting the ket onto the $\vec{x}$ basis, the wavefunction 
$\psi^{(0)}_{a} (t,\vec{x}) \equiv \langle\vec{x}|\nu^{({\rm p},0)}_a (t)\rangle$
satisfies the equation
\begin{equation}\label{Schr0}
-i\frac{\partial\psi_a^{(0)}(t,\vec{x})}{\partial t}=(H_0)_{ab}\psi^{(0)}_b(t,\vec{x})\,,
\end{equation}
with the Hamiltonian $H_0 = \bar{V}\cdot\text{diag}(E_1,..,E_{N_f})\cdot \bar{V}^{-1}$.
The state $\vert\nu^{({\rm p},0)}_a(t_{\rm d})\rangle$ at the moment of detection can be found by solving \cref{Schr0}, which 
accurately describes the dynamics of the superposition of states in \cref{WavePacket} as long as $m_k\ll Q$ and $\sigma\ll m_k^2/Q$, where $Q=(p_{\rm p}+p_{\rm d})/2$ and $k=1,...,N_f$. 
Finally, by integrating $|\mathcal{A}^{(0)}_{ab}|^2$ over the apriori unknown production time $t_p$ and average momentum $p_p$, we obtain the transition probability
\begin{equation}\label{Prob0}
\begin{aligned}
    \mathcal{P}^{(0)}_{ab} & = \limitint_{-\infty}^{\infty}\frac{\diff p_p}{2\pi}\diff t_p\,|\mathcal{A}^{(0)}_{ab}|^2\;, \\   
    & = \sum_{k,l=1}^{N_f} V_{b k}\bar{V}_{a k}\bar{V}_{b l}V_{a l}\: \e^{2\pi i L/L_{kl}^{\rm osc}} \;,
\end{aligned}
\end{equation}
where $L\equiv z_{\rm d}-z_{\rm p}$ is the baseline and $L_{kl}^{\rm osc} = 4\pi Q/\Delta m_{kl}^2$ is the oscillation length, with $\Delta m_{kl}^2=m_k^2-m_l^2$.
In \cref{Prob0}, the effects caused by the dispersion of the wave packets and by their increasing spatial separation are neglected.\:\footnote{The latter effect is often referred to as ``decoherence'' in the neutrino literature. 
We avoid this terminology in order to prevent confusion with the notion of decoherence as maximal entanglement with the environment.}
This is an accurate approximation as long as $L\ll L_{kl}^{\rm coh}$, where $L_{kl}^{\rm coh} = Q^2/(\sigma\Delta m_{kl}^2)$ is the baseline at which the distance between the centers of the wave packets becomes bigger than their spatial width $1/\sigma$.
Furthermore, we neglect the measurement uncertainty of the detector, which is accurate provided that $\sigma\gtrsim 1/L_{kl}^{\rm osc}$.

\section{Nonlinear correction to neutrino propagation}
\label{sec:nonlin}

We would like to modify the oscillation amplitude by introducing a nonlinearity in the neutrino propagation, $\mathcal{A}_{ab} = \langle\nu^{({\rm d})}_b |U(t_{\rm p},t_{\rm d},\nu^{({\rm p})})\nu^{({\rm p})}_a(t_{\rm p})\rangle$.
The nonlinearity manifests itself in the explicit dependence of the modified evolution operator $U$ on the evolving state, represented by $\nu^{(\rm p)}$.
Assuming the correction to the linear evolution amplitude accumulated between the moments of production and detection is small, then one can use perturbation theory with $U(t_{\rm p},t,\nu^{({\rm p})}) = U_0(t_{\rm p},t) + \epsilon\, U_1(t_{\rm p},t,\nu^{({\rm p})})$, where $\epsilon$ is a dimensionless expansion parameter.
Similarly, $|\nu_a^{({\rm p})}(t)\rangle = |\nu_a^{({\rm p},0)}(t)\rangle + \epsilon\, |\nu_a^{({\rm p},1)}(t)\rangle$, where
\begin{equation}
    |\nu_a^{({\rm p},1)}(t)\rangle = U_1(t_{\rm p},t,\nu^{({\rm p},0)}) |\nu_a^{({\rm p})}(t_{\rm p})\rangle \;.  \label{EvolOps2}
\end{equation}
For the correction $\mathcal{P}^{(1)}_{ab}$ to the transition probability (\ref{Prob0}) this implies
\begin{equation}\label{P1}
    \mathcal{P}^{(1)}_{ab} =\frac{1}{\pi} \Re\limitint_{-\infty}^\infty\diff p_{\rm p}\diff t_{\rm p}  \langle  \nu_a^{({\rm p},0)}(t_{\rm d}) | \nu_b^{({\rm d})} \rangle  \langle \nu_b^{({\rm d})} | \nu_a^{({\rm p},1)}(t_{\rm d})\rangle \;.
\end{equation}

To explicitly compute the effect of nonlinearity, we require an equation governing the evolution of the wavefunction $\psi^{(1)}_{a} (t,\vec{x}) = \langle \vec{x} | \nu^{({\rm p},1)}_a (t) \rangle$.
From \cref{EvolOps2} we deduce that 
\begin{equation}\label{Schr1}
    -i\frac{\partial \psi^{(1)}_{a}(t,\vec{x})}{\partial t} = (H_0)_{ab}\psi^{(1)}_{b} (t,\vec{x}) + \mathcal{G}_{a} (t,\vec{x},\psi^{(0)}) \;,
\end{equation}
where the inhomogeneous term $\mathcal{G}_{a}(t,\vec{x},\psi^{(0)})$ contains information about $\psi^{(0)}$ at all times between $t_{\rm p}$ and $t$.
To determine this term, we adopt the framework proposed in \cite{Kaplan:2021qpv} and further studied in \cite{Raizen:2022xlv,Polkovnikov:2022hkg,Broz:2022aea}.
The nonlinearity in this framework arises from promoting interaction terms (involving the neutrino, in our case) in the Lagrangian of the relativistic quantum field theory to state-dependent interactions as in \cref{NLterm}.
For concreteness, let us consider the Yukawa interaction of the form (see Appendix \ref{app:A} for the discussion of the Standard Model neutral current interaction)
\begin{equation}\label{LagrInt}
    \L_{int}= -\frac{v }{2\Lambda_R} Z_{ab}\vf \chi_a^\dagger i\sigma_2\bar{\chi}_b + \text{h.c.} 
\end{equation}
which can be obtained from the Weinberg operator in the electroweak symmetry-broken phase, and which describes the interaction between the Higgs field $\vf$ and the left-handed light active neutrino species at energies much below $\Lambda_R$.
Here $v\simeq 246$ GeV is the Higgs vacuum expectation value, and $\chi_{a}$ denotes the two-component Weyl spinor of the flavour $a$ active neutrino.
The complex, symmetric matrix $Z$ is given by $Z=V\cdot\text{diag}(m_1,...,m_{N_f})\cdot V^T\cdot \Lambda_R/v^2$, where $m_i$ are the neutrino mass eigenstates, and to leading order in $v/\Lambda_R$, the matrix $V$ coincides with the PMNS matrix. 
Next, we add the following modification of the interaction (\ref{LagrInt})
\begin{equation}\label{LagrIntNL}
    \delta\L_{int}=-\epsilon\frac{v }{2\Lambda_R} Z_{ab}\langle\Phi |\hat{\vf}|\Phi\rangle \chi_a^\dagger i\sigma_2\bar{\chi}_b + \text{h.c.}\,,
\end{equation}
where $\epsilon$ is a small dimensionless parameter, $|\Phi\rangle$ is the normalised state of the system, and $\hat{\vf}$ represents the field operator in the Heisenberg picture.
The term (\ref{LagrIntNL}) leads to the nonlinear and nonlocal contribution to the  equation of motion for $\chi_a$.
We write this equation in the form of the Schr\"odinger equation describing the dynamics of relativistic neutrino wave packets, see Appendix \ref{app:B} for more details.
Adopting the ansatz $\chi_a=(\psi_a,\bar{\psi}_a)^T$ and expanding in powers of $\epsilon$, we arrive at \cref{Schr1} where
\begin{equation}\label{InhomTerm}
\begin{split}
    &\mathcal{G}_{a} (t,\vec{x},\psi^{(0)}) =  
    -\frac{1}{v^2}
    \int \diff t'\diff^3 \vec{x}' \bigl\{ G_R(t',\vec{x}',t,\vec{x}) \: m_i m_j \bigr. \\
    &  \bigl. \times\Re[V_{ci}V_{di}]\psi^{(0)}_c(t',\vec{x}')\bar{\psi}^{(0)}_d(t',\vec{x}')\bigr\}V_{aj}V_{bj}\psi^{(0)}_b(t,\vec{x}) \;,
\end{split}
\end{equation}
and $G_R$ is the relativistic retarded Green's function of the massive real scalar field $\vf$. 
The summation runs over $i,j,b,c,d$ from $1$ to $N_f$.

To compute the nonlinear correction, 
we assume that no knowledge of the system history
is required prior to the moment of neutrino production, $\psi_a^{(1)}(t_{\rm p},\vec{x})=0$.
Solving \cref{Schr1,InhomTerm} with this initial condition, one obtains the probability correction (see Appendix \ref{app:C} for the details of this computation)
\begin{equation}\label{P1_final}
\begin{split}
    \mathcal{P}^{(1)}_{a b} = \frac{32\pi^2\gamma_E }{\sqrt{5}v^2} \sum_{i,j,c=1}^{N_f} m_i \Re\left\{ V_{ci}V_{cj}\:\e^{2\pi i L /L_{ij}^{\rm osc}} \right\} & \\
    \times  \sum_{k,l=1}^{N_f} m_k \Im\left\{ V_{ak}V_{bk} V_{al}\bar{V}_{bl}\:\e^{2\pi i L /L_{kl}^{\rm osc}}\right\} & \;.
\end{split}
\end{equation}
This result is derived under the same conditions as the linear vacuum oscillation probability (\ref{Prob0}), and is valid for baselines satisfying $L\ll L_{kl}^{\rm coh}$.
In particular, the conditions $m_k\ll Q$, $\sigma\ll m_k^2/Q$, $k=1,...,N_f$, allow us to remain near the simple plane-wave picture of neutrino oscillations, even though the integration in \cref{InhomTerm} is performed over the neutrino world-line and, thus, requires the neutrino to be localised in space.
Furthermore, in deriving \cref{P1_final} we assumed that the Higgs mass is much larger than the momentum uncertainty $\sigma$, which simplifies the calculation of the integrals in \cref{InhomTerm}.

The nonlinear nature of the correction is manifested in \cref{P1_final} as a product of six PMNS mixing matrix elements as opposed to the four matrix elements in the linear probability $\mathcal{P}^{(0)}_{ab}$.
The presence of an additional $L$-dependent, exponential phase factor in $\mathcal{P}^{(1)}_{ab}$, as compared to $\mathcal{P}^{(0)}_{ab}$, indicates an enhanced interference between the wave packets in the neutrino state $|\nu_a^{(\rm{p})}\rangle$.

Furthermore, the expression (\ref{P1_final}) separates into the product of two terms.
The first term results from the spacetime integral in \cref{InhomTerm} and does not change the flavour of the propagating neutrino.
In fact, in the absence of CP-violation (when $V_{ai}$ is real), this term simply reduces to the sum of neutrino masses.
The second term reflects oscillations induced by the nonlinear correction.
It comes from the fact that the remaining, non-integrated part of \cref{InhomTerm} contains the mixture of all neutrino flavours.
Importantly, the linear and nonlinear-induced oscillations have the same oscillation lengths $L_{ik}^{\rm{osc}}$.
This differs from oscillations, induced, for example, by heavy neutrino states which would be resolved at much shorter baselines.
Note also that the $L$-dependence of $\mathcal{P}^{(1)}_{ab}$ is contained in the exponential phase factors only. This means that when $L\ll L_{kl}^{\rm coh}$, the overall magnitude of the nonlinearity is bounded by a constant. In other words, the nonlinear interference of mass eigenstates does not destabilise the neutrino state $|\nu_a^{(\rm{p})}\rangle$.

The computation in Appendix \ref{app:C} indicates that in the limit of large spatial separation of the mass eigenstates, $L\gg L_{kl}^{\rm coh}$, the nonlinear interference effect vanishes (together with the linear one).
In Appendix~B we argue that the derivation of the nonlinear correction is unchanged if the non-disturbing interaction of the neutrino with the environment is allowed.
At $L\gg L_{kl}^{\rm coh}$, this interaction may cause decoherence of the mass eigenstates and the latter belong to distinct quasiclassical histories. 
We conclude that in our setup the nonlinear modification does not engender a permanent correlation between different branches of the system.
It only manifests itself at distances at which individual oscillations are resolved, and in the rest of the paper we focus on this case.

\section{Physical implications}
\label{sec:phys}

Consider first a two-flavour model with mixing angle $\theta$ and no CP-violation.
Denote the rescaled dimensionless nonlinearity parameter $\tilde\epsilon = 32\pi^2\gamma_E\,\epsilon \,m_{\nu,\rm sum}^2/(\sqrt{5}v^2)$, where $m_{\nu,\rm sum}=\sum_{i=1}^{N_f} m_i$ is the sum of neutrino masses.
From \cref{Prob0,P1_final} one obtains
\begin{subequations}
\begin{align}
    & \P_{ee}=\P_{\mu\mu}=\cos^4\theta+\sin^4\theta +\frac{1}{2}\sin^22\theta \nonumber\\
    & \qquad  \times\left( \cos\left(\frac{2\pi L}{L_{12}^{\rm osc}}\right) -\frac{\tilde\epsilon }{2 }\frac{\Delta m_{21}^2}{m_{\nu,\rm sum}^2}\sin\left(\frac{2\pi L}{L_{12}^{\rm osc}}  \right)\right) \,,\label{P2f11} \\ 
    & \P_{e\mu}=\P_{\mu e} =\sin^22\theta
    \nonumber\\
    & \qquad \times\
    \left( \sin^2 \left(\frac{\pi L}{L_{12}^{\rm osc}}\right) + \frac{\tilde\epsilon }{4 }\frac{\Delta m_{21}^2}{m_{\nu,\rm sum}^2}\sin\left(\frac{2\pi L}{L_{12}^{\rm osc}} \right)\right) \,.
\end{align}
\end{subequations}
First, we see that when the neutrino masses are equal, the $\epsilon$ correction to the oscillation probability vanishes, as expected, since in this case there are no oscillations and $\mathcal{P}_{ab}=\delta_{ab}$.
Second, we observe that the correction changes the survival probability, $\P_{ee}$ by a small factor oscillating with the baseline $L_{21}^{\rm osc}$, while the transition probability, $\P_{e\mu}$ is modified by exactly the opposite factor.
Thus, $\P_{ee}+\P_{e\mu}=1$ (similarly for $\P_{\mu\mu},\P_{\mu e}$), and we conclude that the nonlinear modification preserves unitarity, in agreement with Ref.~\cite{Kaplan:2021qpv}.
Due to the same oscillation period, the correction just shifts the oscillation curve.
For example, the transition probability $\mathcal{P}_{e\mu}$ now attains its maximum at the baseline $L_{21}^{\rm osc}(1/2 - \tilde\epsilon \Delta m_{21}^2/(4\pi m_{\nu,\rm{sum}}^2))$.
The value of $\mathcal{P}_{e\mu}$ at this maximum coincides with that of $\mathcal{P}^{(0)}_{e\mu}$, to first order in $\epsilon$.

\begin{figure}[t]
    \centering
    \includegraphics[width=0.9\linewidth]{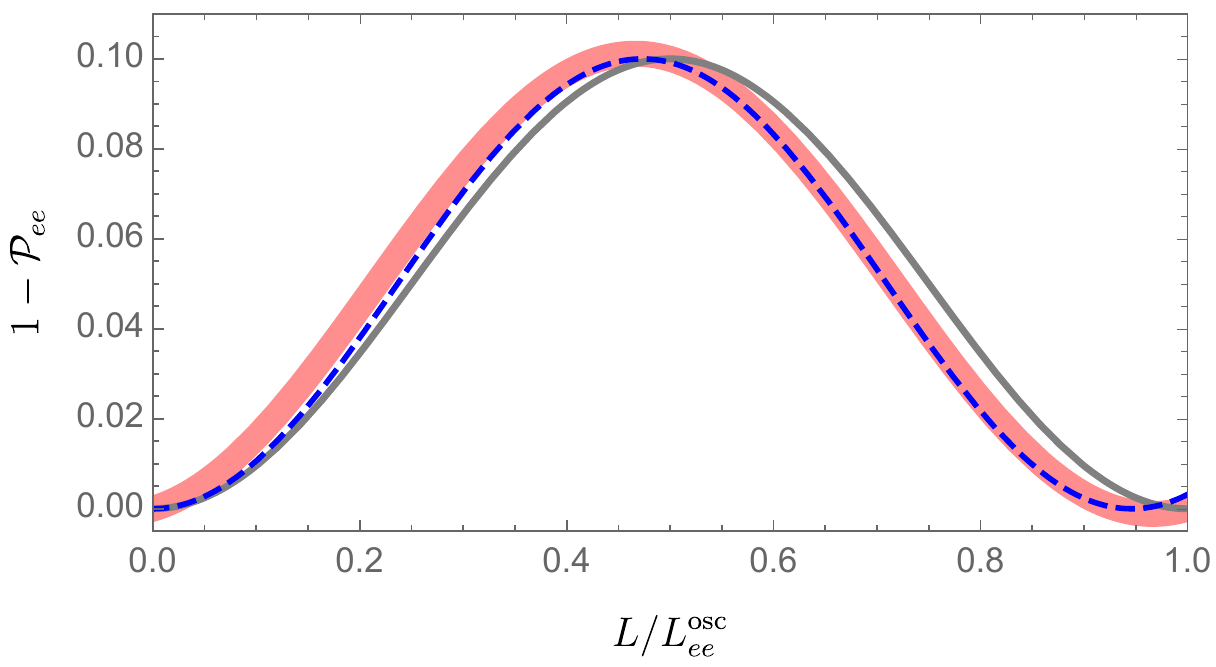}
    \caption{The effect of the nonlinear state-dependent modification due to the Higgs-neutrino interaction on the $\nu_e$ oscillation probability $\mathcal{P}_{ee}$, assuming an effective two-flavour oscillation scheme.
    The oscillation parameters are taken from~\cite{ParticleDataGroup:2020ssz}, and $\tilde{\epsilon}=0.2$.
    The grey solid line shows the original probability $\mathcal{P}_{ee}^{(0)}$.
The red solid band depicts $\mathcal{P}_{ee}$ with $\delta_{CP}=0$ where the thickness represents the  uncertainty from higher powers of $\tilde{\epsilon}$.
The correction induces a shift in the value of $L$ at which the extremum of $\mathcal{P}_{ee}$ is attained.
    The new value 
    can be fit by shifting $L_{ee}^{\rm osc}$ in $\mathcal{P}_{ee}^{(0)}$ (the blue dashed line).
    For $\delta_{CP}\neq 0$, the value of the maximum changes as well, see \cref{EffAngle}.
    }
    \label{fig:plot}
\end{figure}

Next we discuss how the nonlinear correction can impact oscillation data.
Reactor experiments allow for an accurate determination of the mixing angle $\theta_{13}$ via measurements of electron neutrino disappearance. 
In the effective two-flavour oscillation scheme, the survival probability is given by \cite{Nunokawa:2005nx, Minakata:2006gq}
\begin{equation}\label{SurvivalProb}
    \mathcal{P}^{(0)}_{ee}=1-\sin^2 2\theta_{13}\sin^2\left(\frac{\pi L }{L_{ee}^{\rm osc}} \right)\;,
\end{equation}
where $(L_{ee}^{\rm osc})^{-1}=(L_{31}^{\rm osc})^{-1}\cos^2\theta_{12}+(L_{32}^{\rm osc})^{-1}\sin^2\theta_{12}+\mathcal{O}(L_{31}^{\rm osc}/L_{21}^{\rm osc})$.
Further corrections to $L_{ee}$ arise from the heavy Majorana states and are neglected in (\ref{SurvivalProb}).
From \cref{P1_final} we obtain how \cref{SurvivalProb} changes once the nonlinearity is introduced:
\begin{equation}\label{SurvCorr}
\begin{split}
    \epsilon\,\P_{ee}^{(1)} &= \tilde\epsilon\, \frac{\sin^2\theta_{13}}{m_{\nu,\rm sum}}   \left[- m_3 \sin\left(\frac{2\pi L }{L_{ee}^{\rm osc}}+2\delta_{CP}\right)\right.
   \\
   & + \left.\left( m_1\cos^2\theta_{12} +m_2\sin^2\theta_{12}\right) \sin \left(\frac{2\pi L }{L_{ee}^{\rm osc}}\right) \right]\,,
\end{split}
\end{equation}
where $\delta_{CP}$ is the CP-violating phase.
This expression is accurate to leading order in $L_{31}^{\rm osc}/L_{21}^{\rm osc}$ and $\sin^2\theta_{13}$.
We see again that the correction (\ref{SurvCorr}) shifts the value of $L_{ee}^{\rm osc}$ at which the extremum of $\mathcal{P}_{ee}$ is achieved, see Fig.~\ref{fig:plot} for illustration.
Besides, the correction (\ref{SurvCorr}) shifts the measured value $\tilde\theta_{13}$ of the mixing angle $\theta_{13}$:
\begin{equation}\label{EffAngle}
    \sin^2 2\tilde\theta_{13}=\sin^2 2\theta_{13}\left(1-\frac{\tilde\epsilon \,m_3}{4m_{\nu,\rm sum}}\sin \left(2\delta_{CP}\right)\right) \;.
\end{equation}
This is the value measured by terrestrial neutrino experiments with the baselines small compared to the neutrino coherence length.
On the other hand, the value of $\theta_{13}$ inferred from solar neutrino data is not subject to the nonlinear correction, since $L\gg L^{\rm coh}_{31}$ for solar neutrinos.
Taking the experimental values from reactor and solar measurements \cite{ParticleDataGroup:2020ssz}, we obtain the constraint $|\tilde\epsilon|\lesssim 0.3\:\text{eV}/m_3$.
Note that this bound is specifically for the Higgs-neutrino interaction as opposed to bounds on nonlinearity parameters based on electromagnetic interactions~\cite{Raizen:2022xlv,Polkovnikov:2022hkg,Broz:2022aea}.

\section{Discussion}
\label{sec:disc}

The idea that a fundamental nonlinearity exists in the time evolution of an isolated quantum system remains an intriguing possibility. 
The proposal in Ref.~\cite{Kaplan:2021qpv}
provides a prescription to compute the effects of such a nonlinearity. 
In this paper, we have calculated the correction to the oscillation probability that arises from the state-dependent extension of the Yukawa interaction between the light, active neutrino and the Higgs field.
This interaction is proportional to the ratio $v^2/\Lambda_R^2 \sim m^2_{\nu,\rm{sum}}/v^2$, which additionally suppresses the correction (\ref{P1_final}) and makes the resulting bound on the nonlinearity parameter $\epsilon$ not currently relevant for experiment.
Nevertheless, we demonstrated the very existence of the effect induced by the nonlinearity and discussed its general properties following from \cref{P1_final}. 

The Weinberg operator is used as a source of the neutrino-neutrino interaction mediated by the Higgs field.
However, it is straightforward to repeat our analysis for other possible neutrino couplings that would also contribute to the nonlinear interference of the neutrino mass eigenstates and enhance the effect.
For example, in \cref{LagrInt} one can replace $v/\Lambda_R\mapsto y$ and treat $\vf$ as a new scalar field coupled to the neutrino current with the (small) coupling $y$ and the non-diagonal flavour matrix $Z_{ab}$ which is assumed to have order one matrix elements (see, e.g., \cite{Berlin:2016woy}).
The size of the nonlinear correction to the oscillation probability is then proportional to $\epsilon y^2$.
Depending on the nature of the field $\vf$, this can be much larger than (\ref{P1_final}).
Note also that loop effects can potentially transfer the nonlinearity from the scalar-neutrino coupling to different sectors of the Standard Model.
However, we expect such loop-induced contributions to be sub-dominant. 

How is the form of \cref{P1_final} different from corrections induced by the standard neutrino interactions?
One source of correction is the same Yukawa coupling term (\ref{LagrInt}) which, upon integrating out the Higgs field, results in a four-fermion interaction that renormalises the neutrino propagator. 
It is easy to see that this correction scales as $Q^2/\Lambda_R^2$, which differs from the overall scaling $v^2/\Lambda_R^2$ in \cref{P1_final}.
The latter scaling can be reproduced by the one-loop sunrise diagram with the Higgs field.
However, it does not result in the nonlinear interference pattern obtained from the correction (\ref{P1_final}).

Our results are obtained under the assumption of an initial localised neutrino state which may not be possible with realistic mechanisms of neutrino production and will likely require going beyond the wave packet treatment of neutrino propagation.
Nevertheless, the method employed in the neutrino oscillation calculation is quite general and can be used to compute possible consequences of other state-dependent nonlinear interactions in quantum field theory.
Further exploring these fundamental nonlinear effects in particle physics and cosmology opens a way to possibly experimentally probe the many-worlds interpretation of quantum mechanics. 
There should be a world where this is possible, and perhaps even our own.
\\
\\
\noindent{\bf\emph{Acknowledgments.}}---We thank David E. Kaplan, Pedro Machado and Surjeet Rajendran for helpful discussions.
This work is supported in part by the Department of Energy under Grant No. DE-SC0011842. T.G. is also supported by the Simons Foundation.
The work of T.G. was completed at the Aspen Center for Physics, which is supported by National Science Foundation grant PHY-1607611.

\appendix

\section{}
\label{app:A}

The correction  to the oscillation probability studied above is due to the Yukawa interaction (\ref{LagrInt}). 
Here we consider the nonlinear correction from the neutral current interaction
in the Standard Model,
\begin{equation}\label{SM_int}
    \frac{g}{2\cos\theta_W} Z_\mu \sum_{a=1}^3 \chi_a^\dagger\bar\sigma^\mu \chi_a \;,
\end{equation}
where $g$ is the $SU(2)$ gauge coupling and $\theta_W$ is the weak mixing angle.
To determine whether the interaction \eqref{SM_int} gives a nonlinear correction to the oscillation probability we notice that, since the interaction is flavour diagonal, it does not change the oscillation pattern.
Hence, it could only contribute to the total normalisation of the oscillation probability.
This can be checked straightforwardly by repeating the steps described in the Appendix \ref{app:C}.
However, unitarity requires the partial probabilities to always sum up to one.
Thus, at least to first order in $\epsilon$, the correction due to the neutral current interaction must vanish.
To see this more explicitly, consider the relativistic retarded Green's function of the massive vector boson,
\begin{equation}
    G_R^{\mu\nu}(t',\vec{x}',t,\vec{x})=\left( g^{\mu\nu}+\frac{1 }{M_Z^2}\frac{\partial^2 }{\partial x_\mu\partial x_\nu} \right)G_R(t',\vec{x}',t,\vec{x}) \;,
\end{equation}
where $M_Z$ is the $Z$-boson mass, $g^{\mu\nu}$ is the Minkowski metric tensor and $G_R(t',\vec{x}',t,\vec{x})$ is the scalar Green's function.
The corresponding inhomogeneous term is similar to that of \cref{InhomTerm}, and one can write $\mathcal{G}_a(t,\vec{x},\psi^{(0)})\equiv \tilde{\mathcal{G}}(t,\vec{x},\psi^{(0)})\psi_a^{(0)}(t,\vec{x})$ with $\tilde{\mathcal{G}}(t,\vec{x},\psi^{(0)})$ denoting the part in the spacetime integral.
Contracting $G_R^{\mu\nu}$ in this integral with $\bar\sigma^\mu$ and $\bar\sigma^\nu$ from \cref{SM_int} leads to the manifestly real $\tilde{\mathcal{G}}(t,\vec{x},\psi^{(0)})$.
Given that the linear in $\epsilon$ correction to the oscillation probability is proportional to the imaginary part of $\tilde{\mathcal{G}}(t,\vec{x},\psi^{(0)})$ (see \cref{EoM1} below), we conclude that to first order in $\epsilon$, the contribution from the Standard Model neutral current interaction vanishes.

\section{}
\label{app:B}

Here we fill the gaps in the derivation of \cref{InhomTerm}.
To determine how the active neutrino states respond to the background classical field created by the expectation value $\langle\Phi|\hat{\vf}|\Phi\rangle$, we first solve the Klein--Gordon equation for the Higgs field $\vf$ supplemented with the inhomogeneous term (\ref{LagrIntNL}) arising from the interaction (\ref{LagrInt}).\,\footnote{We assume that a proper normalisation procedure has been applied that removed any vacuum divergences in the expectation values.}
The solution can be written as
\begin{equation}\label{phi}
\begin{split}
    \vf(t,\vec{x}) & =-\frac{v}{2\Lambda_R} \int\diff t'\diff\vec{x}' \,G_R(t',\vec{x}',t,\vec{x}) \\
    &\times Z_{ab}\chi_a^\dagger(t',\vec{x}')i\sigma_2\bar{\chi}_b(t',\vec{x}')+\text{h.c.}
\end{split}
\end{equation}
By using the retarded Green's function $G_R$,  the modified theory remains causal \cite{Kaplan:2021qpv}.\footnote{Note that this choice of the Green's function leads to the presence of T-violating terms in the correction to the oscillation probability.}
Next, we promote the fields in \cref{phi} to operators, evaluate the corresponding expectation value in the state $|\Phi\rangle$, and substitute the result into \cref{LagrIntNL}. 
Including the standard spinor kinetic term in the mass basis, $\L_{kin}=i\chi_i^\dagger\bar{\sigma}^\mu\partial_\mu\chi_i$ where $\bar{\sigma}^\mu=(1,-\sigma^i)$, the combined Lagrangian $\L_{kin}+\L_{int}+\delta\L_{int}$, leads to the modified Dirac equation for the flavour neutrino states $\chi_a$.
We are interested in the solutions to this equation which are of the form of the linear superposition of narrow (in momentum space) Gaussian wave packets propagating in vacuum, see eq.~(\ref{WavePacket}). 
For these configurations, the modified Dirac equation reads
\begin{widetext}
\begin{equation}\label{EoM1}
\begin{split}
     -i\frac{\partial\chi_a(t,\vec{x})}{\partial t} &= (H_0)_{ab}\chi_b(t,\vec{x}) + \epsilon\frac{v^2 }{4\Lambda_R^2 }\int\diff t'\diff\vec{x}' G_R(t',\vec{x}',t,\vec{x}) \\
     &\times \langle\Phi| Z_{cd} \hat{\chi}^\dagger_c(t',\vec{x}')i\sigma_2\hat{\bar{\chi}}_d(t',\vec{x}') - \bar{Z}_{dc} \hat{\chi}^T_c(t',\vec{x}')i\sigma_2\hat{\chi}_d(t',\vec{x}') |\Phi\rangle
     Z_{ab}i\sigma_2 \bar{\chi}_b(t,\vec{x}) \;,
\end{split}
\end{equation}
\end{widetext}
where $H_0 = \bar{V}\cdot\text{diag}(E_1,..,E_{N_f})\cdot \bar{V}^{-1}$ and the interaction term \eqref{LagrInt} gives zero contribution for a classical vacuum background.

For a freely propagating neutrino, the state vector of the system is simply $|\Phi\rangle=|\chi_a\rangle$ where the one-particle state $|\chi_a\rangle$ is the eigenvector of $\hat{\chi}_a$.
Hence, we can evaluate the expectation value in the state $|\Phi\rangle$ and replace the operator notation with the wavefunction $\chi$.
This is a good approximation for short enough baselines.
For long distances (e.g., for solar or supernova neutrinos), when the wave packets associated with the neutrino mass eigenstates are widely separated, the interaction with the environment destroys the coherence between the wave packets, and the latter belong to distinct quasiclassical histories. 
Assuming that the interaction does not change appreciably the dynamics of the propagating neutrino, we can account for this decoherence effect by writing $|\Phi\rangle=\bar{V}_{ai}|\chi_i\rangle\otimes |\xi_i\rangle$ where $|\xi_i\rangle$ is the state of the environment (represented, e.g., by a probe particle scattering off one of the wave packets).
Given that in the decoherence limit $|\xi_i\rangle$, $|\xi_j\rangle$ are orthogonal for $i\neq j$, we can evaluate the expectation value in \cref{EoM1} as in the one-particle case.

We apply the ansatz $\chi_a=(\psi_a,\bar{\psi}_a)^T$, which represents equal probabilities for the polarisation of the spinor state (appropriately normalised), where $\psi_a ({\bar\psi}_a)$ are anticommuting variables.
Finally, we expand the wavefunction $\psi_a$ as $\psi_a=\psi_a^{(0)}+\epsilon \psi_a^{(1)}$.
The expansion of \cref{EoM1} to zeroth order in $\epsilon$ is simply the Schr\"odinger \cref{Schr0}.
The expansion to first order in $\epsilon$ gives \cref{Schr1} with the inhomogeneous term 
\begin{equation}\label{G}
\begin{split}
    &\epsilon \,\mathcal{G}_a(t,\vec{x},\psi^{(0)})  \equiv 
-\epsilon \frac{v^2}{\Lambda_R^2}
    \int\diff t'\diff\vec{x}'G_R(t',\vec{x}',t,\vec{x})\\ & \times \Re[Z_{\lbrace cd\rbrace}]
    \psi_c^{(0)}(t',\vec{x}')\bar{\psi}^{(0)}_d(t',\vec{x}') Z_{ab}\psi_b^{(0)}(t,\vec{x}) \;,
\end{split}
\end{equation}
where $Z_{\lbrace cd\rbrace}=(Z_{cd}+Z_{dc})/2$. 
Using that $Z=V\cdot\text{diag}(m_1,...,m_{N_f})\cdot V^T\cdot \Lambda_R/v^2$, we obtain the result (\ref{InhomTerm}).

\section{}
\label{app:C}

Here we derive the nonlinear correction to the neutrino oscillation probability (\ref{P1_final}). 
Our conventions are as follows. 
The delta-function in momentum space satisfies 
\begin{equation}
    \int\diff^3\vec{p}\:\delta^{(3)}(\vec{p}) = 1 \;.
\end{equation}
The completeness of the coordinate eigenstates reads 
\begin{equation}
    \int\diff^3\vec{x}\:\vert\vec{x}\rangle\langle \vec{x}\vert = 1 \;.
\end{equation}
Finally, the neutrino mass eigenstates are normalised as 
\begin{equation}
    \langle \nu_j (\vec{p})\vert \nu_k(\vec{q})\rangle = \delta_{jk}\delta^{(3)}(\vec{p}-\vec{q}) \;,
\end{equation}
where $\delta_{jk}$ is the Kronecker delta.

As explained in the main text, we work under the condition
\begin{equation}\label{Approx1}
\sigma\ll \frac{m_k^2}{Q} \;, ~~ m_k\ll Q \;, ~~ k=1,..,N_f \;,
\end{equation}
where $Q=(p_{\rm p}+p_{\rm d})/2$.
This condition ensures the applicability of eq.~(\ref{Schr0}) to describe the propagation of the superposition of wave packets (\ref{WavePacket}).
Next, we require
\begin{equation}\label{Approx2}
    L\ll L_{kl}^{\rm coh} \;,
\end{equation}
where $L_{kl}^{\rm coh} = Q^2/(\sigma\Delta m_{kl}^2)$ is the baseline at which the distance between the centers of the wave packets becomes bigger than their spatial width $1/\sigma$.
Finally, we assume that 
\begin{equation}\label{Approx3}
    \sigma\gtrsim 1/L_{kl}^{\rm osc} \;,
\end{equation}
where $L_{kl}^{\rm osc} = 4\pi Q/\Delta m_{kl}^2$ is the oscillation length.
Note that the condition (\ref{Approx3}) is compatible with the condition (\ref{Approx1}) provided $m_k^2\gtrsim 10^{-3}$ eV$^2$.

Now we specify the relativistic retarded Green's function of the massive scalar field. 
Assuming $t-t' > |\vec{x}-\vec{x}'| \geqslant 0$, it is given by
\begin{equation}\label{RelGreen}
    G_R(t',\vec{x}',t,\vec{x}) = -\frac{1 }{2\pi}\delta^{(2)}(s^2) + \theta(s^2)\frac{M }{4\pi s}J_1(Ms) \;,
\end{equation}
and vanishes otherwise. 
Here $M$ is the mass of the scalar, $s^2$ is the spacetime interval 
\begin{equation}
    s^2 = (t'-t)^2-\vert \vec{x}'-\vec{x} \vert^2 \;,
\end{equation}
$\theta$ is the Heaviside step-function and $J_1$ is the Bessel function of the first kind.

Next, the solution of eq.~(\ref{Schr1}) can be written as
\begin{equation}\label{psi1}
    \psi_{a}^{(1)}(t,\vec{x}) = \limitint_{t_{\rm p}}^{t_{\rm d}}\diff t'' K_{a b}(t'',t)\mathcal{G}_{b}(t'',\vec{x}) \;,
\end{equation}
where $K_{ab}(t'',t)$ is the Green's function of the Schr\"odinger equation, 
\begin{equation}\label{K}
    \left( -i\delta_{ab}\frac{\diff }{\diff t} - (H_0)_{ab} \right)K_{bc}(t'',t) = \delta_{ac}\delta(t''-t) \;.
\end{equation}
The boundary condition $ \psi_a^{(1)}(t_p,\vec{x})=0 $ suggests that one should take the retarded Green's function which is given by 
\begin{equation}\label{GreenSchr}
    K(t'',t)=i \bar{V} \cdot \text{diag}\:(\e^{-iE_1(t''-t)},...)\cdot \bar{V}^{-1} \cdot \theta (t-t'') \;,
\end{equation} 
as can be checked by substituting \eqref{GreenSchr} into \eqref{K}.

All the necessary ingredients are now in place to compute the correction to the amplitude,
\begin{equation}\label{A1psi}
    \mathcal{A}^{(1)}_{ab} = \langle\nu_b^{({\rm d})} \mid \psi^{(1)}_a(t_{\rm d})\rangle = \int \diff^3 \vec{x} \: \psi_{b}^{({\rm d})\dagger}(\vec{x})\psi_{a}^{(1)}(t_{\rm d},\vec{x}) \;,
\end{equation}
where $\psi_{b}^{({\rm d})}(\vec{x}) \equiv \langle \vec{x}|\nu_b^{({\rm d})}\rangle$.
Consider first the wavefunction squared in the integrand of (\ref{InhomTerm}), which we quote here again for convenience:
\begin{equation}\label{InhomTermApp}
\begin{split}
    &\mathcal{G}_{a} (t,\vec{x},\psi^{(0)}) =  
    -\frac{1}{v^2}
    \int \diff t'\diff^3 \vec{x}' \bigl\{ G_R(t',\vec{x}',t,\vec{x}) \: m_i m_j \bigr. \\
    &  \bigl. \times\Re[V_{ci}V_{di}]\psi^{(0)}_c(t',\vec{x}')\bar{\psi}^{(0)}_d(t',\vec{x}')\bigr\}V_{aj}V_{bj}\psi^{(0)}_b(t,\vec{x}) \;.
\end{split}
\end{equation}
Under the conditions (\ref{Approx1}) and (\ref{Approx2}) we obtain 
\begin{equation}\label{Expr1}
\begin{split}
   & \sum_{i,c,d=1}^{N_f} m_i \Re( V_{ci}V_{di})  \psi_{c}^{(0)}(t',\vec{x}') \bar{\psi}_{d}^{(0)}(t',\vec{x}') = 16\sqrt{2} \pi^{3/2} \sigma^3 \\
    & \times \sum_{i,j,k,c,d=1}^{N_f} m_i \Re( V_{ci}V_{di}) \bar{V}_{c k}V_{d j} \exp\left[ -\frac{i \Delta m_{kj}^2}{2p_{\rm p}}(t_{\rm p}-t') \right]  \\
      & \times \exp\left[ -2\sigma^2((t_{\rm p}-t'-z_{\rm p}+z')^2 +x'^2+y'^2) \right] \;.
\end{split}
\end{equation}
The result \eqref{Expr1} is then substituted into \cref{InhomTermApp} where it is convenient to keep the coordinate and momentum integrals in the expression for $\psi_b^{(0)}(t,\vec{x})$. 
This is followed by substituting \cref{InhomTermApp,GreenSchr} into \cref{psi1}, and finally substituting \cref{psi1} into \cref{A1psi}.
The resulting multiple integrals are then successively evaluated.

First, we integrate over $t'$ in \cref{InhomTermApp} and assume that $M \gg \sigma$ and $M\gg\Delta m_{ij}^2/p_{\rm p}$ for any $i,j=1,..,N_f$.
This allows the $t'$-dependence of the exponent in \cref{Expr1} to be neglected and $t'=t+|\vec{x}'-\vec{x}|$ to be substituted for the lower bound of the integral.
The first term in \cref{RelGreen} gives zero upon integrating over $t'$ and subsequently over $|\vec{x}'-\vec{x}|$.
The second term can be evaluated as follows:
\begin{equation}
    \limitint_{t+|\vec{x}'-\vec{x}|}^\infty \diff t'\; \frac{1}{s} J_1(Ms)= \frac{1 }{M|\vec{x}'-\vec{x}|} \;,
\end{equation}
assuming that $M|\vec{x}'-\vec{x}|\gg 1$. 
This assumption is valid provided the subsequent integration over $|\vec{x}'-\vec{x}|$ is saturated at values much larger than $M^{-1}$.
Next, we change to cylindrical coordinates,
\begin{equation}
      x'-x = r\cos\theta \;, ~~~ y'-y=r\sin\theta \;, ~~~ z'-z=\bar{z}\;,
\end{equation}
so that $|\vec{x}'-\vec{x}|=\sqrt{\bar{z}^2+r^2}$.
One can immediately integrate over $\theta$:
\begin{equation}
    \limitint_0^{2\pi}\diff\theta\:\e^{-4\sigma^2(xr\cos\theta+yr\sin\theta)}=2\pi I_0(4r\sigma^2\sqrt{x^2+y^2}) \;,
\end{equation}
where $I_0$ is the modified Bessel function of the first kind.
Next, we integrate over $t''$ in \cref{psi1} and define
\begin{equation}
    {\cal I}_t = \limitint_{t_{\rm p}}^{t_{\rm d}}\diff t''\e^{iB t''-2\sigma^2 (A-t'')^2} \;,
\end{equation}
where 
\begin{align}
    & A = t_{\rm p}+\bar{z}+z-z_{\rm p}-\sqrt{\bar{z}^2+r^2} \;, \\
    & B = E_n(p)-E_i(Q)+\frac{\Delta m_{kj}^2}{2p_{\rm p}} \;.
\end{align}
Denote $T\equiv t_{\rm d}-t_{\rm p}$ and assume that $T\gg p_{\rm p}/\Delta m_{kj}^2$ for any $k,j=1,..,N_f$ and $T\gg\sigma^{-1}$.
The first assumption is justified in view of the condition (\ref{Approx3}) and the fact that the integral over $p_{\rm p}$ in \cref{P1} is saturated at $p_{\rm p}\approx Q$.
The second assumption is justified for $T\gg L_{kj}^{\rm osc}$ and given \cref{Approx3}, i.e., for baseline distances large compared to the shortest oscillation length.
The limits of integration in ${\cal I}_t$, can then be extended to $t_{\rm p}\to-\infty$, $t_{\rm d}\to+\infty$, provided that the maximum of the exponent, where the integral is saturated, is located between $t_{\rm p}$ and $t_{\rm d}$, namely
\begin{equation}\label{Cond2}
    t_{\rm p}< A <t_{\rm d} \;.
\end{equation}
Evaluating the integral, we obtain
\begin{equation}
    {\cal I}_t = \sqrt{\frac{\pi }{2\sigma^2}}\e^{iAB-\frac{B^2 }{8\sigma^2 }} \;.
\end{equation}
Next, we integrate over $\bar{z}$ and define
\begin{equation}
    {\cal I}_z=\limitint_{\bar{z}_{\rm p}}^{\bar{z}_{\rm d}}\diff\bar{z}\frac{1 }{\sqrt{\bar{z}^2+r^2}}\e^{\frac{i\Delta m_{kj}^2}{2p_{\rm p}}\sqrt{\bar{z}^2+r^2}+iB\left( \bar{z}-\sqrt{\bar{z}^2+r^2} \right)} \;,
\end{equation}
where the limits of integration must obey the condition (\ref{Cond2}).
It is easy to see that if $z<z_{\rm p}$, \cref{Cond2} is never satisfied, hence the integral is zero. 
Next, if $z_{\rm p}<z<z_{\rm p}+T$, the lower limit is finite but the upper limit is infinite,
\begin{equation}\label{limits2}
    \bar{z}_{\rm p}=\frac{(z_{\rm p}-z)^2-r^2}{2(z_{\rm p}-z)} \;,~~~ \bar{z}_{\rm d}=\infty \;.
\end{equation}
Finally, if $z>z_{\rm p}+T$, both limits are finite.
Let us focus on the second case as the physical one. 
Note that the integral over $z$ in \cref{A1psi} is saturated around $z_{\rm d}\approx z_{\rm p}+T$, since the detector wave packet is concentrated around $z=z_{\rm d}$.
Hence, one can substitute $z=z_{\rm d}$ in \cref{limits2}. 
Assuming that the integral over $r$ is saturated at $r\ll L$, this gives $\bar{z}_{\rm p}\approx -L/2$, which, in turn, can be safely replaced by $\bar{z}_{\rm p}\to-\infty$.
Thus, we obtain
\begin{equation}
    {\cal I}_z=\limitint_{-\infty}^{\infty}\diff\bar{z}\frac{1 }{\sqrt{\bar{z}^2+r^2}}\e^{iC\left( \sqrt{\bar{z}^2+r^2}-\bar{z} \right)+iD \bar{z}} \;,
\end{equation}
where
\begin{equation}
    C = E_i(Q)-E_n(p) \;, ~~~ D = \frac{\Delta m_{kj}^2}{2p_{\rm p}} \;.
\end{equation}
Assuming that the integral over $r$ is saturated at $r\ll C^{-1}$, gives
\begin{equation}\label{I_x_final}
    {\cal I}_z=-2\gamma_E-\log\left(\frac{D}{4}(2C-D)r^2\right) \;,
\end{equation}
where $\gamma_E \approx 0.577$ is the Euler--Mascheroni constant.
Next, we integrate over $r$:
\begin{equation}\label{I_r}
    \limitint_0^\infty\diff r \, r\e^{-2\sigma^2r^2}{\cal I}_z I_0(4r\sigma^2\sqrt{x^2+y^2})=-\frac{\gamma_E}{2\sigma^2}\e^{2\sigma^2(x^2+y^2)} \;,
\end{equation}
where the logarithm in \cref{I_x_final} is neglected, since under the assumptions $C,D\ll\sigma$, it never becomes significant.
The latter assumptions are equivalent to the condition (\ref{Approx3}) given that the subsequent integration over $p$ picks up the value $p\approx p_{\rm p}$ and the integration over $p_{\rm p}$ in \cref{P1} picks up the value $p_{\rm p}\approx Q$. 
We also see that the integral (\ref{I_r}) is saturated at $r\sim\sigma^{-1}$, justifying the assumptions made in evaluating ${\cal I}_z$ (provided the condition (\ref{Approx3}) is valid).

It remains to integrate over $\vec{x}$ in \cref{A1psi} using the explicit expression (\ref{WavePacket}) for the wave packets.
The integral over $\vec{x}$ then produces a delta function that removes one of the momentum integrals. 
The remaining momentum integral is straightforward. 
Assuming that 
\begin{equation}\label{Approx5}
    \frac{L\sigma^2 }{Q}\ll 1\,,
\end{equation}
gives the expression 
\begin{widetext}
\begin{equation}\label{A1}
\begin{split}
&   \mathcal{A}^{(1)}_{ab}=  i\frac{8\pi^2\gamma_E }{\sqrt{3}v^2} \sum_{c,k,j=1}^{N_f} m_k \Re \left\{ V_{ck} V_{cj} \exp\left[ \frac{i \Delta m_{kj}^2 L }{2Q} \right] \right\}  \sum_{d,i,n=1}^{N_f} m_n \bar{V}_{ai}V_{di}V_{dn}V_{bn} \exp\left[ \frac{i m_i^2 T}{2Q} \right]  \\
  &\qquad\qquad\qquad\qquad\qquad\qquad\times \exp\left[- \frac{i\Delta m_{in}^2 L}{2Q} -\frac{(p_{\rm p}-p_{\rm d})^2 }{12\sigma^2}+iQ(T-L) \right] \;.
\end{split}
\end{equation}
The correction to the transition probability is given by 
\begin{equation}
    \mathcal{P}^{(1)}_{ab} = \frac{1}{\pi} \Re\limitint_{-\infty}^\infty \diff p_{\rm p}\diff t_{\rm p}\, \mathcal{A}^{(0)}_{ab}\bar{\mathcal{A}}^{(1)}_{ab} \;,
\end{equation}
and the integration is straightforward. 
The final result is 
\begin{equation}\label{P1}
  \mathcal{P}^{(1)}_{ab} =\frac{32\pi^2\gamma_E }{\sqrt{5}v^2} \sum_{c,k,j=1}^{N_f} m_k \Re \left\lbrace V_{ck} V_{cj} \exp\left[ \frac{2\pi i L }{L_{kj}^{\rm osc}} \right] \right\rbrace  \sum_{d,i,n,l=1}^{N_f} m_n \Im\left\lbrace\bar{V}_{ai}V_{di}V_{dn}V_{bn}V_{al}\bar{V}_{bl} 
     \exp\left[ \frac{2\pi i L }{L_{nl}^{\rm osc}} \right] \right\rbrace \exp\left[- \frac{2\pi^2}{(\sigma L_{il}^{\rm osc})^2} \right] \;.
\end{equation}
\end{widetext}
The expression simplifies in the regime (\ref{Approx3}), and we obtain (\ref{P1_final}).

Under the condition (\ref{Approx3}), the assumption (\ref{Approx5}) leads to (\ref{Approx2}).
Thus, \cref{P1} (or (\ref{P1_final})) describes the nonlinear interference between the mass eigenstates of the propagating neutrino.
The opposite limit, $L\sigma^2/Q\gg 1$, follows again from \cref{Approx3} and $L\gg L_{kl}^{\rm coh}$.
In this case the wave packets in the superposition (\ref{WavePacket}) are widely separated and, as discussed in Appendix~B, can be assumed to have been decohered.
Evaluating the momentum integral in this regime, one finds that the resulting amplitude is suppressed relative to (\ref{A1}) by the factor $Q/(L \sigma^2 )$. Thus, we conclude that at large baselines the nonlinear effect is gradually washed out.

\bibliography{Refs}

\end{document}